\documentclass[pre,amsmath,amssymb,letterpaper,twocolumn,10pt]{revtex4}
\pdfoutput=1

\usepackage{amsmath}
\usepackage{hyperref}
\usepackage{graphicx}
\usepackage{natbib}

\usepackage{setspace}
\usepackage{soul}

\usepackage[margin=2cm]{geometry}
\def\kt{k_{\rm B}T}

\def\f#1{Fig.~\ref{#1}}

\def\beq{\begin{equation}}
\def\eeq{\end{equation}}
\def\bea{\begin{eqnarray}}
\def\eea{\end{eqnarray}}

\usepackage{color}
\def\boxit#1{\vbox{\hrule\hbox{\vrule\kern6pt\vbox{\kern6pt#1\kern6pt}\kern6pt\vrule}\hrule}}

\begin{document}
\title{Selective nucleation in porous media}
\author{Lester O. Hedges}
\author{Stephen Whitelam}
\affiliation{Molecular Foundry, Lawrence Berkeley National Laboratory, Berkeley, CA 94720, USA}

\begin{abstract}
Geometrical arguments suggest that pore-mediated nucleation happens in general in a two-step fashion, the first step being nucleation within the pore, the second being nucleation from the filled pore into solution [Page \& Sear, Phys. Rev. Lett. 97, 65701 (2006)]. The free energy barriers controlling the two steps of this process show opposite dependencies on pore size, implying that for given thermodynamic conditions there exists a pore size for which nucleation happens fastest. Here we show, within the two- and three-dimensional Ising lattice gas, that this preferred pore size tracks the size of the bulk critical nucleus, up to a numerical prefactor. This observation suggests a simple prescription for directing nucleation to certain locations within heterogeneous porous media.
\end{abstract}

\maketitle

\section{Introduction}
\vspace{-1.0em}

Nucleation of a new phase can be faster on a flat surface than in the bulk, provided that the surface possesses sufficient attraction for the new phase~\cite{Young:1805,Turnbull:1950,winter2009monte}. Nucleation within a pore made of the same material can be faster still, because pore corners provide energetically preferred binding sites at which the new phase can take hold~\cite{page2006heterogeneous}. Further, for given thermodynamic conditions, arguments of geometry alone suggest a pore size for which nucleation is fastest~\cite{page2006heterogeneous}. Pore-mediated nucleation involves, in general, two free energy barriers, one confronting nucleation within the pore, the second confronting nucleation from the filled pore into solution (see \f{fig1} inset). These barriers show opposing dependencies on pore size: the `in' barrier gets bigger with increasing pore size, because more unfavorable surface must be created by the new phase to span the pore; the `out' barrier gets smaller with increasing pore size, because the new phase filling the pore presents a larger interface to solution~\cite{page2006heterogeneous,hedges2012patterning}. As a result, nucleation into solution is fastest, for given thermodynamic conditions, for a given pore size.

Here we illustrate the flip side of this argument: as one varies thermodynamic conditions, one changes the pore size for which nucleation is fastest. As stated in Ref.~\cite{page2006heterogeneous}, the preferred pore size should be about the size of the bulk critical nucleus, because a pore much larger or much smaller influences nucleation not much differently than does a flat surface. Here we confirm this statement within the 2D and 3D Ising lattice gas, showing that one can control the hierarchy of nucleation rates mediated by a set of pores of different sizes by controlling supersaturation. This control occurs because the preferred pore size tracks the size of the bulk critical nucleus, up to a numerical prefactor, and so therefore scales as the reciprocal of the bulk driving force for nucleation. This observation suggests a simple strategy for directing nucleation to specified locations within heterogeneous porous media, both natural and artificial.

\section{Model and simulation methods}
We simulated nucleation in the Ising lattice gas in two and three dimensions~\cite{stauffer1982monte,brendel2005nucleation,acharyya1998nucleation,wonczak2000confirmation,binder1974investigation,Maibaum:2008}, on square and cubic lattices respectively, in the presence of walls. The model energy function is
\beq
E=-J \sum_{\langle ij\rangle }n_i n_j - \Delta \mu \sum_i n_i - J_{\rm{s}}\sum^{\rm wall}_{ij} n_i n_j^{\rm w}.
\eeq
Here $n_i=0$ denotes a vacancy at site $i$, while $n_i=1$ denotes a particle at that location. $J$, the nearest-neighbor coupling, sets the surface tension between particle- and vacancy phases. $\Delta \mu$ is the chemical potential (hereafter `supersaturation') that can be tuned to favor particles or vacancies. The first sum runs over all distinct nearest-neighbor bulk bonds, and the second sum runs over all bulk sites. The third term, whose sum runs over all bonds connecting bulk sites and wall sites, describes interactions between particles and walls sites $(n_j^{\rm w} =1)$. In what follows we set $ J=3.2 \, \kt$ and 1.6 (2D and 3D, respectively) corresponding to $55$\% of the Ising critical temperature, and $J_{\rm s}=1.6 \, \kt$ and 0.8 (in 2D and 3D). Our results are qualitatively insensitive to variation of $J_{\rm s}$, provided that it is large enough to render nucleation faster on a surface than in the bulk~\cite{hedges2012patterning}.

We carried out simulations using a standard grand canonical Metropolis Monte Carlo algorithm~\cite{frenkel1996understanding}, in concert with umbrella sampling~\cite{torrie1977nonphysical,Pan:2004,wolde1999homogeneous} and unbiasing methods~\cite{Ferrenberg:1989,Kastner:2005} to determine free energy landscapes for nucleation, and forward-flux sampling~\cite{Allen:2006} to determine rates of nucleation. Simulation box sizes were $100^2$ lattice sites in 2D and $30^3$ lattice sites in 3D. Pores of fixed depth were used throughout (30 sites in 2D, 10 sites in 3D). More details of the methods used are given in Ref.~\cite{hedges2012patterning}. As in that paper, we note that our simulation scheme ignores effects of mass transport and effects like the mismatch in registry between a new phase and its template~\cite{van2010design}. The effects identified here are instead generic features of geometry.

\section{Results}

In \f{fig1} we show that the likelihood of nucleation in and out of a pore of a given width (in 2D) varies with supersaturation. On the vertical axis we plot $R_w/\sum_{w'} R_{w'}$, where $R_w = (1/R_w^{\rm in} + 1/R_w^{\rm out})^{-1}$ is the mean rate of nucleation in and out of a pore of width $w$~\cite{page2006heterogeneous}. The sum runs over all 6 widths shown, namely $14,12,10,8,6$ and 0 lattice sites, the latter meaning a flat surface. Two regimes are evident. At large supersaturation, nucleation rates are similar for all pores considered (all are larger than on a flat wall made of the same material). This regime is one of low `selectivity', in the sense that one cannot distinguish between different pores on the basis of their associated nucleation rates. At smaller supersaturation, however, the `selective' regime identified in Ref.~\cite{page2006heterogeneous} is found. In this regime, pore nucleation rates become distinct. At certain supersaturations, nucleation is much more likely to happen in the presence of some pores than others, and as supersaturation varies, each pore in turn can be made more likely than its competitors to give rise to a nucleation event.
\begin{figure}[t!]
\center
\includegraphics[width=\linewidth]{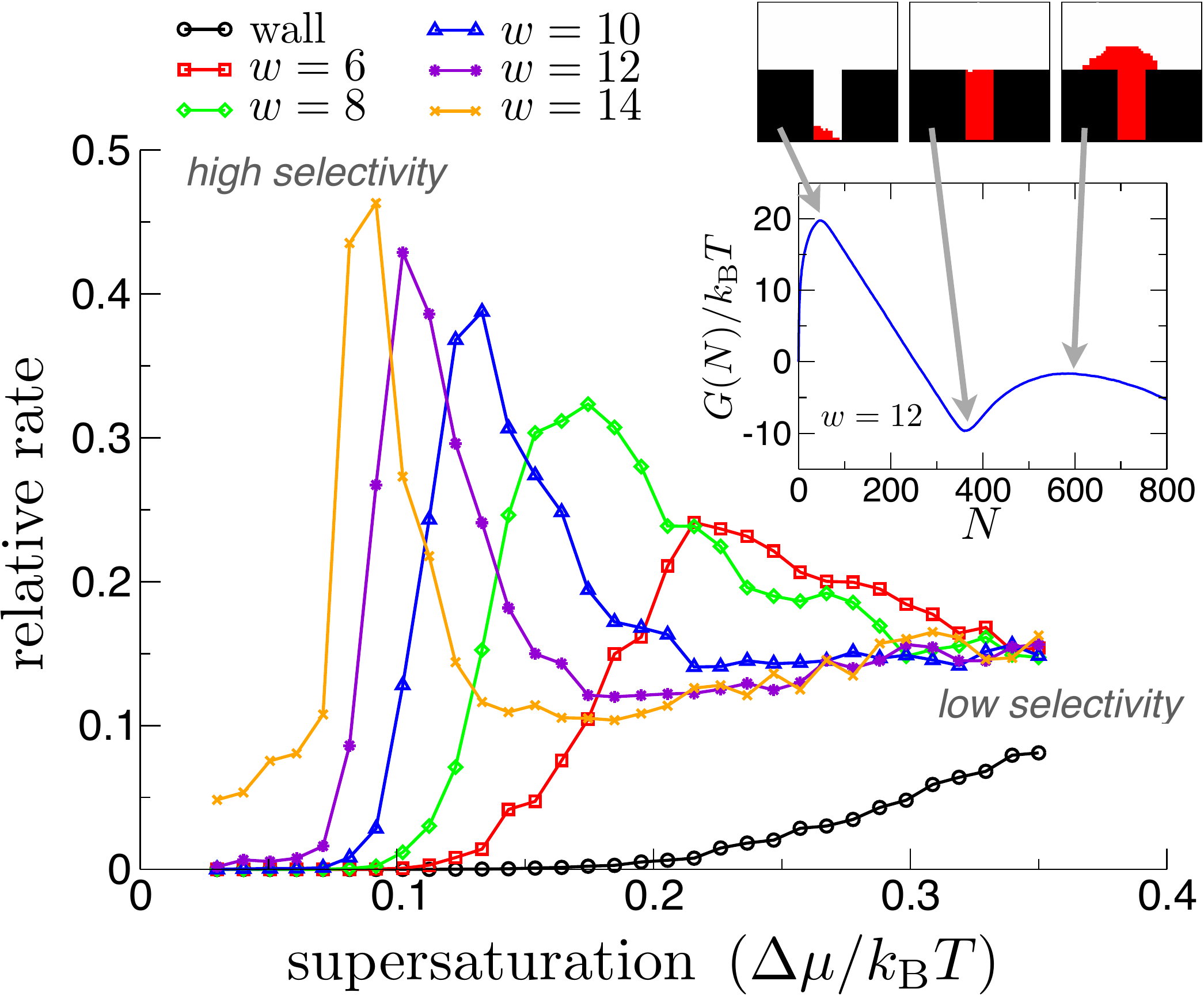}
\caption{{\em Relative nucleation rates within a set of pores change with supersaturation}. We plot relative rates of nucleation mediated by 5 different pores of widths $w$ (and by a flat wall), as a function of supersaturation. At small supersaturation a `selective' regime is encountered, in which nucleation is more likely in the presence of some pores than others. Inset: illustration of the double free energy barrier for nucleation in and out of a pore, as a function of the number of particles $N$ in the largest cluster. The resulting two-step nucleation mechanism~\cite{page2006heterogeneous} makes possible the selection mechanism shown in the main figure.}
\label{fig1}
\end{figure}

In \f{fig2} we show that within a set of pores, different hierarchies of nucleation rates can be obtained with different choices of supersaturation. At the larger of the two supersaturations shown, nucleation into solution happens fastest in the presence of a pore of width 9 lattice sites. Given three pores of widths 15, 12 and 9 lattice sites, nucleation happens first in the presence of the smallest pore with likelihood 55\%, happens first in the presence of the intermediate pore with likelihood 28\%, and happens first in the presence of the largest pore with likelihood 17\%. For the smaller of the two supersaturations shown, however, this hierarchy is reversed: nucleation happens first in the presence of the pore of width 9 with essentially zero likelihood. Instead, it happens first in in the presence of the pores of width 15 and 12 sites with likelihood 90\% and 10\%, respectively. Thus by changing supersaturation, one can select, in a statistical sense, which pore will mediate nucleation into solution. 
\begin{figure}[b!]
\center
\includegraphics[width=\linewidth]{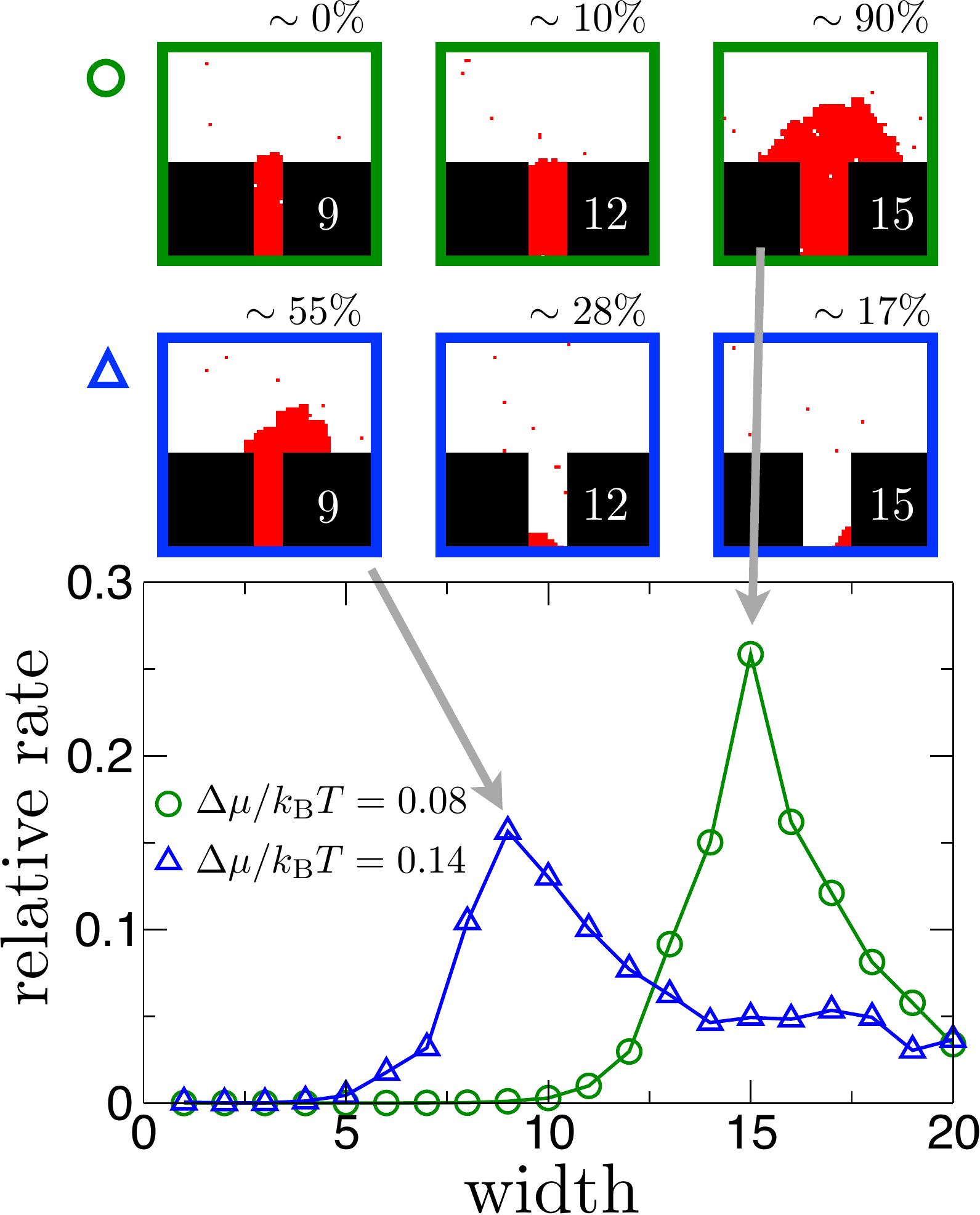}
\caption{{\em Within a set of pores, different hierarchies of nucleation rates can be obtained with different choices of supersaturation.} The hierarchy of nucleation rates mediated by set of three pores (pictured top) can be reversed by changing supersaturation. Numbers next to each snapshot show the likelihood with which nucleation in and out of that pore happens first. Configurations are shown at a time characteristic of nucleation out of the preferred pore. The bottom panel shows the relative rates of nucleation mediated by a larger set of pores at the same two supersaturations, emphasizing that the preferred pore size changes with supersaturation. (Note that the percentages given above snapshots assume a competition between three pores, while the relative rates shown in the bottom panel assume a competition between 20 pores.)}
\label{fig2}
\end{figure}
\begin{figure*}[t!]
\center
\includegraphics[width=\linewidth]{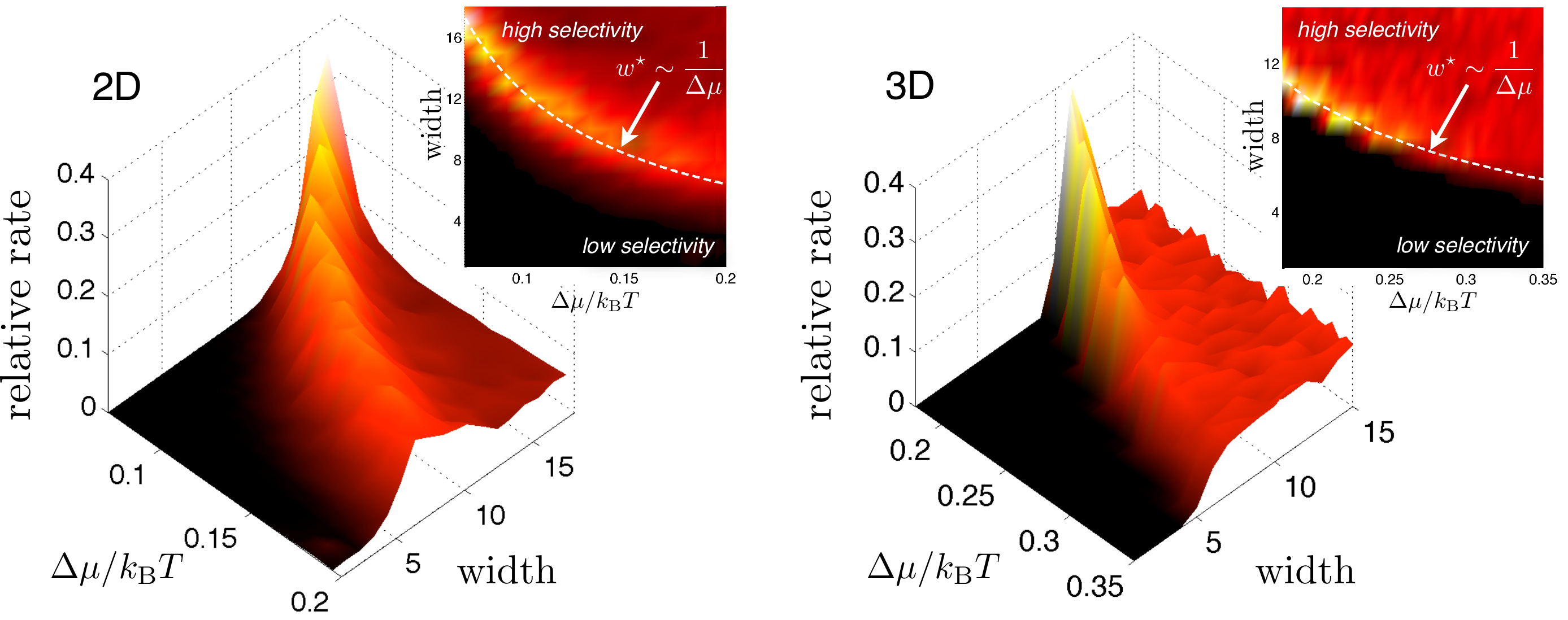}
\caption{{\em The optimal pore size tracks the size of the bulk critical nucleus in two and three dimensions}. We show relative nucleation rate as a function of pore size and supersaturation, overlaid (insets) by dotted lines denoting 100\% and $140$\% of the radius of the respective bulk critical nucleus.  In both 2D and 3D (in the `selective' nucleation regime) the pore that mediates nucleation fastest falls on this dotted line. The preferred pore size in both cases therefore scales, at small supersaturations, as the reciprocal of the bulk driving force, $w^{\star}\sim 1/\Delta \mu$. This scaling suggests a simple prescription for directing nucleation to certain locations within heterogeneous porous media.}
\label{fig3}
\end{figure*}

This selection mechanism varies strongly with supersaturation. In \f{fig3} (left) we show relative rates of nucleation mediated by a range of pores of widths 0 (a flat surface) to 20, as a function of supersaturation. Relative rates $R_w/\sum_{w'} R_{w'}$ were computed at each supersaturation; note that the {\em absolute} rate of nucleation summed across all pore sizes (not shown) simply increases with supersaturation. The sharp spike shows that the pore in whose presence nucleation is fastest increases in size with decreasing supersaturation. As stated in Ref.~\cite{page2006heterogeneous}, the preferred pore size tracks the size of the bulk critical nucleus: if a pore is much larger or much smaller than the bulk critical nucleus then it influences nucleation not much differently than does a flat surface. In the inset we illustrate this statement by showing a plan view of the main figure. The radius of the bulk critical nucleus is overlaid as a dotted line (we obtained this radius analytically, by maximizing Eq. (3) of Ref.~\cite{hedges2012patterning}, and assuming a spherical nucleus). The preferred pore size therefore scales as $w^{\star} \sim 1/\Delta \mu $ at small supersaturation.

In \f{fig3} (right) we show that the same physics holds in three dimensions. The main panel shows relative rates for nucleation in the presence of square pores of given edge length $w$. The dotted line in the inset is $140$\% of the bulk critical nucleus radius (we determined the latter by umbrella sampling). Again, therefore, the preferred pore size $w^{\star}$ scales as $\sim 1/\Delta \mu $ at small supersaturation.

\section{Conclusions}
A considerable experimental literature~\cite{gelb1999phase} shows that pores can strongly influence phase equilibria and phase change. We have illustrated, within the 2D and 3D Ising lattice gas, one of the conclusions stated in Ref.~\cite{page2006heterogeneous}: geometry alone suggests that the pore width best for promoting nucleation into solution, $w^{\star}$, is about the size of the bulk critical nucleus, $r_{\rm b}$. At $55$\% of the critical temperatures of these models we found $w^{\star} \approx r_{\rm b}$ (2D) and $w^{\star} \approx 1.4 r_{\rm b}$ (3D). These results suggest that given a set of pores of different sizes one can dictate, in a statistical sense, which pore is most likely to promote nucleation. Such control would offer a way of directing nucleation to specified locations within heterogeneous porous media, such as in natural rock formations or on nanopatterned substrates. 

However, it should be borne in mind that our simulations ignore several effects that may be important in real physical systems. First, because we have used a simple lattice model, our pore surfaces have no molecular detail. Experiments show that such detail can be important: for instance, pore surface chemistry strongly affects the rate of aspirin nucleation~\cite{diao2011surface}. Second, because we have used a simple lattice model, the nucleating phase itself has no internal structure, and so we cannot consider selection by a pore between different polymorphs of the nucleating phase. But such selection does happen: for instance, certain pores prefer certain crystal structures of ice to be nucleated within them~\cite{baker1997nucleation}. In addition, the internal structure of a nucleus can be crucial in determining the rate of nucleation of a new phase~\cite{shah2011computer}. To capture such detail, one would need to use an off-lattice simulation model~\cite{van2010design}. Second, our simulation protocol, grand-canonical Monte Carlo, ignores effects of mass transport that may be important for real pores: our pores cannot become blocked, for instance, which may be a concern when considering viscous fluids in rock pores. To address those effects requires numerical methods designed to treat hydrodynamic flow~\cite{sahimi2012flow}.

That said, the virtue of using a simple model is that it allows one to identify generic effects -- here of geometry -- that may be common to physical systems {\em unlike} in chemical and molecular detail. Geometry alone dictates that there exists a free energy barrier to nucleation in 2D and 3D space; such a barrier is indeed seen in the case of a wide variety of real systems~\cite{sear2007nucleation}. The preferred pore size scaling behavior, identified in Ref.~\cite{page2006heterogeneous}, is also a consequence of geometry alone, and it is this mechanism that makes possible the selective nucleation in pores of different sizes that we have demonstrated within a simple model system. It is therefore possible that this effect could be observed, and made use of, in a wide variety of real systems.

\section{Acknowledgements}
L.O.H. was supported by the Center for Nanoscale Control of Geologic CO$_2$, a U.S. D.O.E. Energy Frontier Research Center, under Contract No. DE-AC02--05CH11231. This work was done at the Molecular Foundry, Lawrence Berkeley National Laboratory, supported under Contract No. DE-AC02-05CH11231. This research used resources of the National Energy Research Scientific Computing Center, which is supported by the Office of Science of the U.S. Department of Energy under Contract No. DE-AC02-05CH11231.


\begin{thebibliography}{25}
\expandafter\ifx\csname natexlab\endcsname\relax\def\natexlab#1{#1}\fi
\expandafter\ifx\csname bibnamefont\endcsname\relax
  \def\bibnamefont#1{#1}\fi
\expandafter\ifx\csname bibfnamefont\endcsname\relax
  \def\bibfnamefont#1{#1}\fi
\expandafter\ifx\csname citenamefont\endcsname\relax
  \def\citenamefont#1{#1}\fi
\expandafter\ifx\csname url\endcsname\relax
  \def\url#1{\texttt{#1}}\fi
\expandafter\ifx\csname urlprefix\endcsname\relax\def\urlprefix{URL }\fi
\providecommand{\bibinfo}[2]{#2}
\providecommand{\eprint}[2][]{\url{#2}}

\bibitem[{\citenamefont{Young}(1805)}]{Young:1805}
\bibinfo{author}{\bibfnamefont{T.}~\bibnamefont{Young}},
  \bibinfo{journal}{Philosophical Transactions of the Royal Society of London}
  \textbf{\bibinfo{volume}{95}}, \bibinfo{pages}{65} (\bibinfo{year}{1805}).

\bibitem[{\citenamefont{Turnbull}(1950)}]{Turnbull:1950}
\bibinfo{author}{\bibfnamefont{D.}~\bibnamefont{Turnbull}},
  \bibinfo{journal}{Journal of Applied Physics} \textbf{\bibinfo{volume}{21}},
  \bibinfo{pages}{1022} (\bibinfo{year}{1950}).

\bibitem[{\citenamefont{Winter et~al.}(2009)\citenamefont{Winter, Virnau, and
  Binder}}]{winter2009monte}
\bibinfo{author}{\bibfnamefont{D.}~\bibnamefont{Winter}},
  \bibinfo{author}{\bibfnamefont{P.}~\bibnamefont{Virnau}}, \bibnamefont{and}
  \bibinfo{author}{\bibfnamefont{K.}~\bibnamefont{Binder}},
  \bibinfo{journal}{Physical Review Letters} \textbf{\bibinfo{volume}{103}},
  \bibinfo{pages}{225703} (\bibinfo{year}{2009}).

\bibitem[{\citenamefont{Page and Sear}(2006)}]{page2006heterogeneous}
\bibinfo{author}{\bibfnamefont{A.}~\bibnamefont{Page}} \bibnamefont{and}
  \bibinfo{author}{\bibfnamefont{R.}~\bibnamefont{Sear}},
  \bibinfo{journal}{Physical Review Letters} \textbf{\bibinfo{volume}{97}},
  \bibinfo{pages}{65701} (\bibinfo{year}{2006}).

\bibitem[{\citenamefont{Hedges and Whitelam}(2012)}]{hedges2012patterning}
\bibinfo{author}{\bibfnamefont{L.~O.} \bibnamefont{Hedges}} \bibnamefont{and}
  \bibinfo{author}{\bibfnamefont{S.}~\bibnamefont{Whitelam}},
  \bibinfo{journal}{Soft Matter} \textbf{\bibinfo{volume}{8}},
  \bibinfo{pages}{8624} (\bibinfo{year}{2012}).

\bibitem[{\citenamefont{Stauffer et~al.}(1982)\citenamefont{Stauffer, Coniglio,
  and Heermann}}]{stauffer1982monte}
\bibinfo{author}{\bibfnamefont{D.}~\bibnamefont{Stauffer}},
  \bibinfo{author}{\bibfnamefont{A.}~\bibnamefont{Coniglio}}, \bibnamefont{and}
  \bibinfo{author}{\bibfnamefont{D.}~\bibnamefont{Heermann}},
  \bibinfo{journal}{Physical Review Letters} \textbf{\bibinfo{volume}{49}},
  \bibinfo{pages}{1299} (\bibinfo{year}{1982}).

\bibitem[{\citenamefont{Brendel et~al.}(2005)\citenamefont{Brendel, Barkema,
  and van Beijeren}}]{brendel2005nucleation}
\bibinfo{author}{\bibfnamefont{K.}~\bibnamefont{Brendel}},
  \bibinfo{author}{\bibfnamefont{G.}~\bibnamefont{Barkema}}, \bibnamefont{and}
  \bibinfo{author}{\bibfnamefont{H.}~\bibnamefont{van Beijeren}},
  \bibinfo{journal}{Physical Review E} \textbf{\bibinfo{volume}{71}},
  \bibinfo{pages}{031601} (\bibinfo{year}{2005}).

\bibitem[{\citenamefont{Acharyya and Stauffer}(1998)}]{acharyya1998nucleation}
\bibinfo{author}{\bibfnamefont{M.}~\bibnamefont{Acharyya}} \bibnamefont{and}
  \bibinfo{author}{\bibfnamefont{D.}~\bibnamefont{Stauffer}},
  \bibinfo{journal}{The European Physical Journal B-Condensed Matter and
  Complex Systems} \textbf{\bibinfo{volume}{5}}, \bibinfo{pages}{571}
  (\bibinfo{year}{1998}).

\bibitem[{\citenamefont{Wonczak et~al.}(2000)\citenamefont{Wonczak, Strey, and
  Stauffer}}]{wonczak2000confirmation}
\bibinfo{author}{\bibfnamefont{S.}~\bibnamefont{Wonczak}},
  \bibinfo{author}{\bibfnamefont{R.}~\bibnamefont{Strey}}, \bibnamefont{and}
  \bibinfo{author}{\bibfnamefont{D.}~\bibnamefont{Stauffer}},
  \bibinfo{journal}{The Journal of Chemical Physics}
  \textbf{\bibinfo{volume}{113}}, \bibinfo{pages}{1976} (\bibinfo{year}{2000}).

\bibitem[{\citenamefont{Binder and
  M{\"u}ller-Krumbhaar}(1974)}]{binder1974investigation}
\bibinfo{author}{\bibfnamefont{K.}~\bibnamefont{Binder}} \bibnamefont{and}
  \bibinfo{author}{\bibfnamefont{H.}~\bibnamefont{M{\"u}ller-Krumbhaar}},
  \bibinfo{journal}{Physical Review B} \textbf{\bibinfo{volume}{9}},
  \bibinfo{pages}{2328} (\bibinfo{year}{1974}).

\bibitem[{\citenamefont{Maibaum}(2008)}]{Maibaum:2008}
\bibinfo{author}{\bibfnamefont{L.}~\bibnamefont{Maibaum}},
  \bibinfo{journal}{Phys. Rev. Lett.} \textbf{\bibinfo{volume}{101}},
  \bibinfo{pages}{256102} (\bibinfo{year}{2008}).

\bibitem[{\citenamefont{Frenkel and Smit}(1996)}]{frenkel1996understanding}
\bibinfo{author}{\bibfnamefont{D.}~\bibnamefont{Frenkel}} \bibnamefont{and}
  \bibinfo{author}{\bibfnamefont{B.}~\bibnamefont{Smit}},
  \emph{\bibinfo{title}{{Understanding Molecular Simulation: From Algorithms to
  Applications}}} (\bibinfo{publisher}{Academic Press, Inc. Orlando, FL, USA},
  \bibinfo{year}{1996}), ISBN \bibinfo{isbn}{0122673700}.

\bibitem[{\citenamefont{Torrie and Valleau}(1977)}]{torrie1977nonphysical}
\bibinfo{author}{\bibfnamefont{G.}~\bibnamefont{Torrie}} \bibnamefont{and}
  \bibinfo{author}{\bibfnamefont{J.}~\bibnamefont{Valleau}},
  \bibinfo{journal}{Journal of Computational Physics}
  \textbf{\bibinfo{volume}{23}}, \bibinfo{pages}{187} (\bibinfo{year}{1977}).

\bibitem[{\citenamefont{Pan and Chandler}(2004)}]{Pan:2004}
\bibinfo{author}{\bibfnamefont{A.}~\bibnamefont{Pan}} \bibnamefont{and}
  \bibinfo{author}{\bibfnamefont{D.}~\bibnamefont{Chandler}},
  \bibinfo{journal}{J. Phys. Chem. B} \textbf{\bibinfo{volume}{108}},
  \bibinfo{pages}{19681} (\bibinfo{year}{2004}).

\bibitem[{\citenamefont{Wolde and Frenkel}(1999)}]{wolde1999homogeneous}
\bibinfo{author}{\bibfnamefont{P.~R.} \bibnamefont{Wolde}} \bibnamefont{and}
  \bibinfo{author}{\bibfnamefont{D.}~\bibnamefont{Frenkel}},
  \bibinfo{journal}{Physical Chemistry Chemical Physics}
  \textbf{\bibinfo{volume}{1}}, \bibinfo{pages}{2191} (\bibinfo{year}{1999}).

\bibitem[{\citenamefont{Ferrenberg and Swendsen}(1989)}]{Ferrenberg:1989}
\bibinfo{author}{\bibfnamefont{A.~M.} \bibnamefont{Ferrenberg}}
  \bibnamefont{and} \bibinfo{author}{\bibfnamefont{R.~H.}
  \bibnamefont{Swendsen}}, \bibinfo{journal}{Phys. Rev. Lett.}
  \textbf{\bibinfo{volume}{63}}, \bibinfo{pages}{1195} (\bibinfo{year}{1989}).

\bibitem[{\citenamefont{K{\"a}stner and Thiel}(2005)}]{Kastner:2005}
\bibinfo{author}{\bibfnamefont{J.}~\bibnamefont{K{\"a}stner}} \bibnamefont{and}
  \bibinfo{author}{\bibfnamefont{W.}~\bibnamefont{Thiel}},
  \bibinfo{journal}{The Journal of Chemical Physics}
  \textbf{\bibinfo{volume}{123}}, \bibinfo{pages}{144104}
  (\bibinfo{year}{2005}).

\bibitem[{\citenamefont{Allen et~al.}(2006)\citenamefont{Allen, Frenkel, and
  ten Wolde}}]{Allen:2006}
\bibinfo{author}{\bibfnamefont{R.~J.} \bibnamefont{Allen}},
  \bibinfo{author}{\bibfnamefont{D.}~\bibnamefont{Frenkel}}, \bibnamefont{and}
  \bibinfo{author}{\bibfnamefont{P.~R.} \bibnamefont{ten Wolde}},
  \bibinfo{journal}{The Journal of Chemical Physics}
  \textbf{\bibinfo{volume}{124}}, \bibinfo{pages}{024102}
  (\bibinfo{year}{2006}).

\bibitem[{\citenamefont{van Meel et~al.}(2010)\citenamefont{van Meel, Sear, and
  Frenkel}}]{van2010design}
\bibinfo{author}{\bibfnamefont{J.}~\bibnamefont{van Meel}},
  \bibinfo{author}{\bibfnamefont{R.}~\bibnamefont{Sear}}, \bibnamefont{and}
  \bibinfo{author}{\bibfnamefont{D.}~\bibnamefont{Frenkel}},
  \bibinfo{journal}{Physical Review Letters} \textbf{\bibinfo{volume}{105}},
  \bibinfo{pages}{205501} (\bibinfo{year}{2010}).

\bibitem[{\citenamefont{Gelb et~al.}(1999)\citenamefont{Gelb, Gubbins,
  Radhakrishnan, and Sliwinska-Bartkowiak}}]{gelb1999phase}
\bibinfo{author}{\bibfnamefont{L.~D.} \bibnamefont{Gelb}},
  \bibinfo{author}{\bibfnamefont{K.}~\bibnamefont{Gubbins}},
  \bibinfo{author}{\bibfnamefont{R.}~\bibnamefont{Radhakrishnan}},
  \bibnamefont{and}
  \bibinfo{author}{\bibfnamefont{M.}~\bibnamefont{Sliwinska-Bartkowiak}},
  \bibinfo{journal}{Reports on Progress in Physics}
  \textbf{\bibinfo{volume}{62}}, \bibinfo{pages}{1573} (\bibinfo{year}{1999}).

\bibitem[{\citenamefont{Diao et~al.}(2011)\citenamefont{Diao, Myerson, Hatton,
  and Trout}}]{diao2011surface}
\bibinfo{author}{\bibfnamefont{Y.}~\bibnamefont{Diao}},
  \bibinfo{author}{\bibfnamefont{A.~S.} \bibnamefont{Myerson}},
  \bibinfo{author}{\bibfnamefont{T.~A.} \bibnamefont{Hatton}},
  \bibnamefont{and} \bibinfo{author}{\bibfnamefont{B.~L.} \bibnamefont{Trout}},
  \bibinfo{journal}{Langmuir} \textbf{\bibinfo{volume}{27}},
  \bibinfo{pages}{5324} (\bibinfo{year}{2011}).

\bibitem[{\citenamefont{Baker et~al.}(1997)\citenamefont{Baker, Dore, and
  Behrens}}]{baker1997nucleation}
\bibinfo{author}{\bibfnamefont{J.}~\bibnamefont{Baker}},
  \bibinfo{author}{\bibfnamefont{J.}~\bibnamefont{Dore}}, \bibnamefont{and}
  \bibinfo{author}{\bibfnamefont{P.}~\bibnamefont{Behrens}},
  \bibinfo{journal}{The Journal of Physical Chemistry B}
  \textbf{\bibinfo{volume}{101}}, \bibinfo{pages}{6226} (\bibinfo{year}{1997}).

\bibitem[{\citenamefont{Shah et~al.}(2011)\citenamefont{Shah, Santiso, and
  Trout}}]{shah2011computer}
\bibinfo{author}{\bibfnamefont{M.}~\bibnamefont{Shah}},
  \bibinfo{author}{\bibfnamefont{E.~E.} \bibnamefont{Santiso}},
  \bibnamefont{and} \bibinfo{author}{\bibfnamefont{B.~L.} \bibnamefont{Trout}},
  \bibinfo{journal}{The Journal of Physical Chemistry B}
  \textbf{\bibinfo{volume}{115}}, \bibinfo{pages}{10400}
  (\bibinfo{year}{2011}).

\bibitem[{\citenamefont{Sahimi}(2011)}]{sahimi2012flow}
\bibinfo{author}{\bibfnamefont{M.}~\bibnamefont{Sahimi}},
  \emph{\bibinfo{title}{Flow and transport in porous media and fractured rock:
  From classical methods to modern approaches, {\em Wiley-Vch, ISBN:
  978-3-527-40485-8}}} (\bibinfo{year}{2011}), \bibinfo{edition}{2nd} ed.

\bibitem[{\citenamefont{Sear}(2007)}]{sear2007nucleation}
\bibinfo{author}{\bibfnamefont{R.~P.} \bibnamefont{Sear}},
  \bibinfo{journal}{Journal of Physics: Condensed Matter}
  \textbf{\bibinfo{volume}{19}}, \bibinfo{pages}{033101}
  (\bibinfo{year}{2007}).

\end{thebibliography}
\end{document}